%% file: manuscript.tex
\documentclass[english,prl,aps,twocolumn,10pt]{revtex4-2}

\usepackage[T1]{fontenc}
\usepackage{amsmath}
\usepackage{amssymb}
\usepackage{graphicx}
\usepackage{babel}
\usepackage{multirow}
\usepackage{array}
\usepackage{verbatim}
\usepackage[colorlinks=true, pdfstartview=FitV, linkcolor=blue, citecolor=blue, urlcolor=blue]{hyperref} 
\usepackage{xcolor}
\usepackage{bm}
\usepackage{tikz}

\def\eg       {{\it e.g.}}
\def\ie       {{\it i.e.}}

\newcommand{\ee}[1]{\cdot10^{#1}}
\newcommand{\mr}[1]{\mathrm{#1}}
\newcommand{\unit}[1]{\,\mathrm{#1}}
\newcommand{\um}{\,\mu{\rm m}}

\newcommand{\uW}{\,\mu{\rm W}}

\newcommand{\oC}{^\circ{\rm C}}

\newcommand{\ye}{\gamma_\mr{e}}

\newcommand{\BNV}{B_\mr{NV}}
\newcommand{\Bo}{B_0}
\newcommand{\Bx}{B_x}

\newcommand{\Bz}{B_z}

\newcommand{\Ileak}{I_\mr{L}}
\newcommand{\Pd}{P_\mr{d}}
\newcommand{\Pgate}{P}

\newcommand{\rd}{r_\mr{d}}
\newcommand{\ro}{r_0}
\newcommand{\vecr}{\bm r}

\newcommand{\Vgate}{V_\mr{G}}
\newcommand{\Vgateon}{V_\mr{G}^\mr{ON}}
\newcommand{\zo}{z_0}

\newcommand{\rhoc}{\rho_\mr{c}}

\begin{document}

\title{Imaging of Gate-Controlled Suppression of Superconductivity\newline via the Meissner Effect}

\author{P.~J.~Scheidegger$^{1\dagger}$, K.~J.~Knapp$^{1\dagger}$, U.~Ognjanovi\'c$^{1\dagger}$, L.~Ruf$^{2}$, S.~Diesch$^{1}$, E.~Scheer$^{2}$, A.~Di~Bernardo$^{2,3}$, and C.~L.~Degen$^{1,4\ast}$}
\affiliation{$^1$Department of Physics, ETH Z\"urich, Otto Stern Weg 1, 8093 Z\"urich, Switzerland.}
\affiliation{$^2$Department of Physics, University of Konstanz, Universit\"atsstrasse 10, 78464 Konstanz, Germany.}
\affiliation{$^3$Dipartimento di Fisica ``E. R. Caianiello'', Universit\`a degli Studi di Salerno, via Giovanni Paolo II 132, 84084 Fisciano, Salerno, Italy.}
\affiliation{$^4$Quantum Center, ETH Z\"urich, 8093 Z\"urich, Switzerland.}
\affiliation{$^\dagger$These authors contributed equally. $^\ast$\href{mailto:degenc@ethz.ch}{degenc@ethz.ch}}

\begin{abstract}
It was recently discovered that supercurrents flowing through thin superconducting nanowires can be quenched by a gate voltage. This gate control of supercurrents, known as the GCS effect, could enable superconducting transistor logic.  Here, we report that the GCS also manifests in a suppression of Meissner screening, establishing the phenomenon as a genuine feature of superconductivity that is not restricted to transport.  Using a scanning nitrogen-vacancy magnetometer at sub-Kelvin temperatures, we image the nanoscale spatial region of GCS suppression in micron-size niobium islands. Our observations are compatible with a microscopic hot-spot model of quasiparticle generation and diffusion, and in conflict with other candidate mechanisms such as Joule heating or an electric field effect. Our work introduces an alternative means for studying quasiparticle dynamics in superconducting nanostructures, and showcases the power of local imaging techniques for understanding emergent condensed matter phenomena.
\end{abstract}
 
\date{\today}
	
\maketitle
		

The recent discovery of a gate-controlled supercurrent (GCS) in superconducting nanoconstrictions~\cite{desimoni2018} has sparked significant scientific interest, owing to the possible new physics and to potential applications in transistors for low power dissipation electronics~\cite{likharev2012,holmes2013,desimoni2021,ryu2024}. A defining feature of the GCS is the complete switching of a typically small superconducting volume into its normal state under the application of a sufficiently large gate voltage.
Multiple mechanisms have been proposed as the physical origin of the GCS~\cite{ruf2024}.  These include a direct $E$-field effect~\cite{desimoni2018,puglia2020,rocci2020,rocci2021}, Joule heating~\cite{catto2022} and quasiparticle excitations in the superconductor.  Quasiparticles, which are formed from broken Cooper pairs~\cite{kivelson1990}, can be excited by electron injection \cite{alegria2021,jalabert2023,basset2021,elalaily2023a} or by high-energy phonons~\cite{ritter2021,ritter2022,elalaily2021} that are created through inelastic scattering of charge carriers in the substrate.  The GCS could also arise from a combination of mechanisms rather than a single process~\cite{ruf2024}.  Although significant work has been put into developing a microscopic understanding of the GCS effect, its precise origin is still under debate.  

While the GCS effect has been studied in a multitude of transport-based experiments with various device geometries, its effect on other properties of superconductivity, such as the Meissner screening of magnetic fields, has not yet been explored.  Moreover, there is a lack of spatial investigation of the GCS beyond the one-dimensional information gained from multi-gate measurements~\cite{desimoni2018,ritter2021}.  A recent scanning tunneling microcopy (STM) experiment varied the lateral position of the STM tip and the tip voltage to study the relation between quasiparticle injection and critical current suppression in superconducting nanowires~\cite{jalabert2023}.


In this work, we report on the spatial imaging of the GCS in Nb island microstructures via the Meissner effect, complementing transport-based experiments and confirming the GCS as a genuine feature affecting superconductivity.  Using scanning NV magnetometry at temperatures between $0.4-5.8\unit{K}$, we demonstrate complete suppression of superconductivity exceeding distances of $1\um$ from the gate electrode.  The suppression is caused by the leakage current between gate and island and depends on the dissipated power, rather than the applied gate voltage.
Our results are consistent with the formation of a local hot spot at the location of highest current density, generating high-energy particles (such as phonons) that radially diffuse through the substrate and excite quasiparticles in the superconductor.  These quasiparticles then locally destroy superconductivity.  The measured diffusion length is $\sim 0.6\unit{\um}$ and the estimated quasiparticle lifetime $\sim 10^{-9}\unit{s}$. We also observe a peculiar interaction between the GCS and superconducting vortices already at very low power.


\subsection*{Superconducting island devices}

\begin{figure}[ht]
	\centering
	\includegraphics[width=0.50\textwidth]{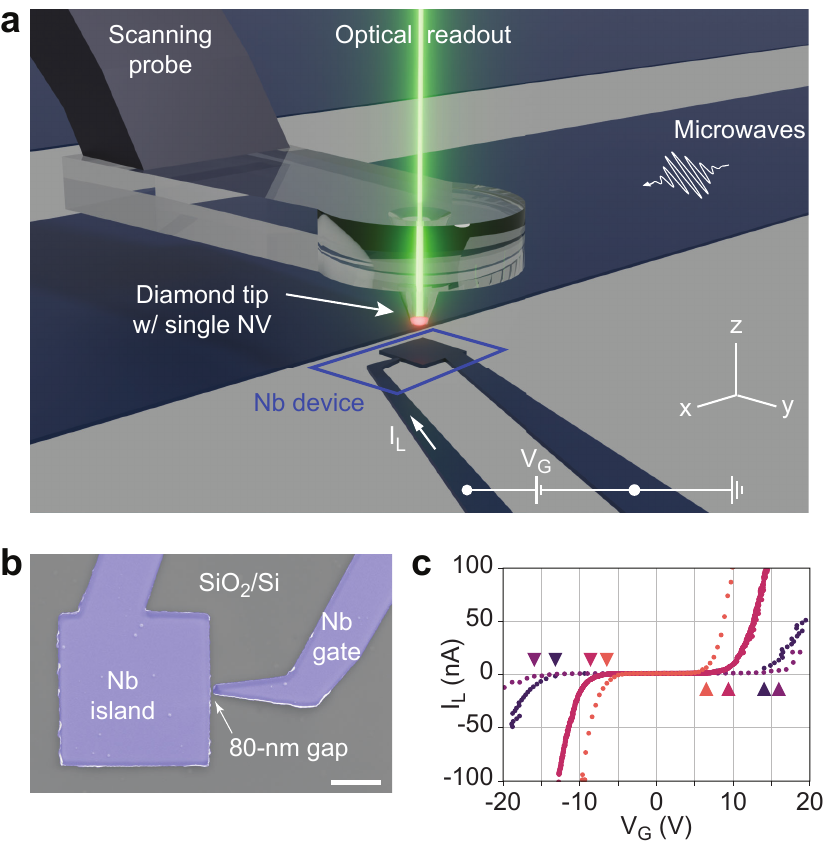}
	\caption{
		\textbf{Experimental setup and superconducting devices.}
		{\bf a}, Sketch of the scanning NV magnetometer and superconducting sample.  We image the stray field above the superconducting Nb device (dark blue) by scanning with the diamond NV probe (red) at constant height ($\zo\sim 100\unit{nm}$).
		{\bf b}, False-colored scanning electron micrograph of a device identical to D2a and D2b. The superconducting Nb structure (blue) sits on top of a SiO$_2$ substrate (gray). A voltage $\Vgate$ is applied between island and gate in constant-current mode (leakage current $\Ileak$). Scale bar, $1\unit{\um}$.
		{\bf c}, $\Ileak/\Vgate$ curves measured on device D2b. Multiple curves (colors) are acquired over time, showing significant variation in the onset voltages of $\Ileak$.  The triangles mark the onset voltages $\Vgateon$ defined by the condition $P = \Ileak\Vgateon = 50\unit{nW}$.
	}
	\label{fig1}
\end{figure}

Our devices consist of approximately 20-nm-thick Nb films on top of a $\sim 5\unit{nm}$ Ti adhesion layer, using intrinsic Si with a $300\unit{nm}$ oxide layer as the substrate.  The devices are patterned using a lift-off process~\cite{ruf2024} (Methods).  We fabricate three geometries in separate fabrication runs: the first geometry (device D1, Fig.~S1) is a nanowire designed to mimic transport-based GCS devices~\cite{desimoni2018}.  The second geometry (devices D2a and D2b, Fig.~\ref{fig1}{\bf b}) consists of a nominally $3\unit{\um}\times 3\unit{\um}$ square island and a finger gate with a $\sim 80\unit{nm}$ gap.  The third geometry (device D3, Fig.~\ref{fig4}{\bf a}) has a significantly larger Nb island with multiple gate geometries.  Each of the devices is placed next to a coplanar waveguide for applying microwave pulses to the NV center magnetic field sensor~\cite{scheidegger2022}.  The superconducting properties measured on device D1 are summarized in Table~S1.

Fig.~\ref{fig1}\textbf{c} shows the leakage-current-to-gate-voltage ($\Ileak/\Vgate$) characteristics for device D2b.  We apply a voltage $\Vgate$ between the gate and island through ohmic leads and measure the resulting leakage current $\Ileak$ (Methods).  We reproduce a sharp increase of $\Ileak$ when reaching an onset voltage $\left|\Vgateon\right| \sim 7-16\unit{V}$ (triangles in Fig.~\ref{fig1}{\bf c}) that is typical for the onset of GCS~\cite{ritter2021}.  The relatively large $\Ileak$ of up to several hundred nA compared to a few nA or less for transport studies~\cite{ruf2024} are attributed to the increased dimensions of our devices as well as to the micron-scale distances over which we probe the Meissner suppression.
Interestingly, we find that $\Vgateon$ fluctuates over time, resembling recent reports of variable stress-induced leakage current in similar devices~\cite{elalaily2024, ruf2024b} (Methods).  In the following, we report our results in terms of the dissipated power, $\Pgate = \Ileak\Vgate$.  We find below that the experimental hallmark of the GCS -- the suppression of superconductivity -- is reproducible given the same applied power $\Pgate$ for all our devices.

\subsection*{Qualitative observations from magnetometry scans}

\begin{figure*}[t]
	\centering
	\includegraphics[width=0.99\textwidth]{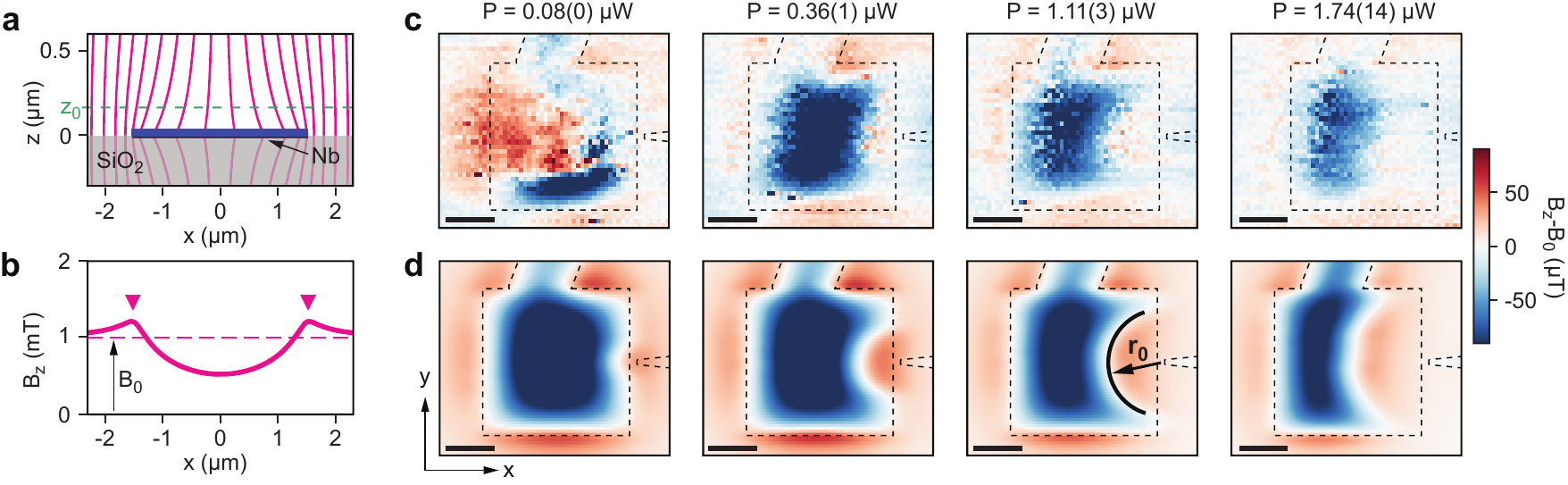}
	\caption{
		\textbf{Observation of gate-controlled suppression of the Meissner effect.}
		{\bf a}, Principle of Meissner screening in an out-of-plane magnetic field $B_0$. The stray field lines (purple) are partially expelled from the thin Nb film (blue). The green dashed line indicates the scan height $\zo$.
		{\bf b}, Model calculation of stray field at height $\zo$. Triangles mark the field maxima that approximately coincide with the edges of the superconducting film. The dashed line is the applied background field $B_0$.  
		{\bf c}, Experimentally measured stray field maps plotted for increasing leakage power $P$.  Blue regions of low field reflect the diamagnetic Meissner screening by the sample.  Note the presence of superconducting vortices in the first panel (red $\Bz$ maxima) that are absent in the later panels taken at higher $P$.  $B_0 = 1\unit{mT}$. 
		{\bf d}, Corresponding simulated magnetic field maps, assuming a circular suppression of superconductivity with radius $\ro$ centered at the gate apex and no vortices. Simulations are performed using the SuperScreen software package~(Ref.~\cite{bishopvanhorn2022} and Methods).  Scale bars in {\bf c,d}, $1\unit{\um}$.
	}
	\label{fig2}
\end{figure*}

To detect superconductivity in the Nb islands, we expose the devices to a small out-of-plane bias field ($\Bo\sim 1\unit{mT}$) and image the magnetic field expulsion due to the Meissner effect.  This results in lower stray field above the center of the islands and higher stray field along the edges (Fig.~\ref{fig2}{\bf a,b})~\cite{rohner2018}.  The stray field maxima roughly delineate the superconducting region.  Suppression of superconductivity due to the GCS effect will then manifest as a spatial shift of the field maximum away from the gate electrode.  We image the stray fields using a custom-built scanning NV magnetometer operating inside a dilution refrigerator~\cite{scheidegger2022} (Fig.~\ref{fig1}a and Methods).

We first report on qualitative measurements performed on device~D2b at $T=0.45\unit{K}$. Fig.~\ref{fig2}{\bf c} shows the reconstructed $\Bz$ component of the measured stray field (Methods) at increasing gate power $\Pgate$.  Except for the lowest power where vortices are present, discussed below, all magnetometry scans show the characteristic reduction of the stray field above the superconducting island.  The gate and ohmic leads are barely resolved because their width is of order of the penetration depth ($\lambda \sim 240\unit{nm}$, Table~S2).  For higher powers, a clear shrinking of the low-field area is observed on the right, gate-facing side of the island.  Qualitatively, the imaged stray fields are in good agreement with the simulations depicted in Fig.~\ref{fig2}{\bf d}.  These observations are clear evidence for a GCS effect in our Nb film and are the first key result of our study.

Curiously, near zero gate power, we observe a cluster of four superconducting vortices within the Nb island (Fig.~\ref{fig2}{\bf c}).  Such vortex formation in applied bias fields is typical for thin-film superconductors ~\cite{embon2017}. Interestingly, the vortices completely vanish from the island at the onset of the GCS, even when the suppressed area is only a small fraction of the total device area (second panel in {\bf c}).  This suggests an interplay of the GCS with vortices, for example, through a modified potential at the superconducting-to-normal boundary or an interaction of quasiparticles with vortices.  We briefly speculate about this feature in the outlook section.

\subsection*{Quantitative analysis of quenching}

Next, we establish a quantitative relationship between the gate power $\Pgate$ and the spatial extent of the quenching of superconductivity.  We develop two methods for quantifying the quenched region; with both methods, we assume that the quenching extends radially from the tip of the gate electrode up to a critical distance $\ro$. This model is further justified below.  In the first approach, we fit horizontal line-cuts of the measured stray field by numerical simulations of superconducting rectangles of varying width~\cite{supplementary}.  In addition to giving an estimate for $\ro$, these fits yield values of important device parameters, such as the superconducting penetration depth (Table~S2).  However, because this method does not account for vortices, it cannot be applied to the entire dataset.  The second approach involves locating the field step in a $\Bx$ map and determining the shift relative to a reference map measured with $\Vgate=0$~\cite{supplementary}.

\begin{figure}[ht]
	\centering
	\includegraphics[width=0.50\textwidth]{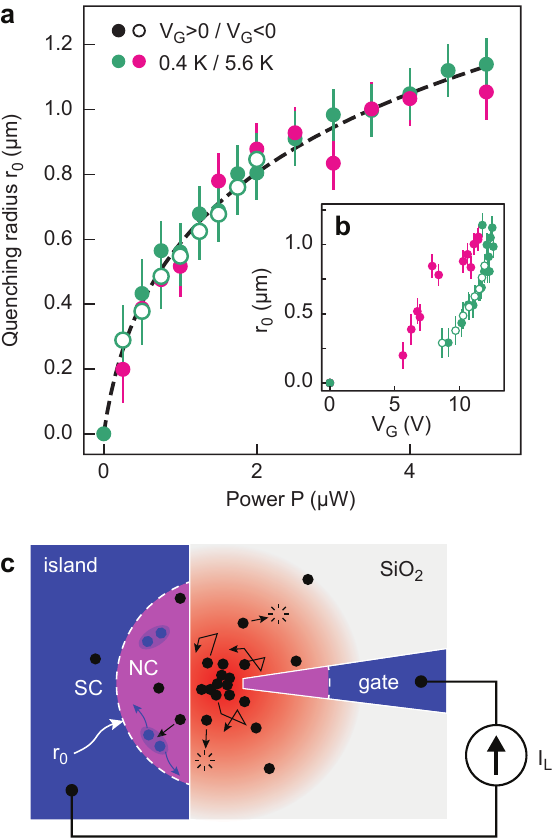}
	\caption{
		\textbf{Power-dependence of GCS effect and hot-spot model of suppressed superconductivity.}
		{\bf a}, Suppression radius $\ro$ plotted as a function of gate power $\Pgate$, for positive and negative gate polarities (open and closed circles, respectively) and two bath temperatures (colors).  The dashed line is a fit to Eq.~(\ref{eq:ro}). Data are from device D3, gate G1.
		{\bf b}, Data of {\bf a} plotted against gate voltage $|\Vgate|$ showing only weak correlation thereby ruling out a direct field effect.
		{\bf c}, Hot-spot model.  Non-equilibrium particles (black dots), generated near the gate apex where the current density is highest, diffuse through the SiO$_2$ substrate.  In the Nb film, they break up Cooper pairs creating quasiparticles (blue dots), turning the superconductor (SC) into a normal conductor (NC) within the critical radius $\ro$.
	}
	\label{fig3}
\end{figure}

The results of this analysis are presented in Fig.~\ref{fig3}{\bf a} for device D3.  The dataset combines measurements at two temperatures ($0.4\unit{K}$, $5.6\unit{K}$) and at positive and negative gate voltages collected during a single cool-down.  Strikingly, the quenching of superconductivity is only dependent on the gate power, and independent of temperature and gate polarity within experimental uncertainty.  Moreover, the quenching of superconductivity extends over large distances exceeding $1\unit{\um}$ at higher gate powers.  Datasets from devices D2a and D2b (Figs.~S3 and S4), which include a broader range of experimental parameters and measurements from several cool-downs, confirm this observation.

Fig.~\ref{fig3} represents the second key result of our study, because the data allow us to rule out several candidate mechanisms for the gate-controlled supercurrent effect in our devices. First, Joule heating caused by the leakage current can be excluded because the quenching does not depend on temperature. Second, electron injection through vacuum is ruled out because the effect does not depend on gate polarity~\cite{elalaily2021,ruf2024}. Finally, a direct electric field effect is also excluded because the quenching is not correlated with the gate voltage (Fig.~\ref{fig3}{\bf b}).


\subsection*{Hot-spot model}

In the following, we argue that the quenching of superconductivity is caused by non-equilibrium particles (NEPs) -- namely, hot electrons or phonons~\cite{ritter2021,ritter2022,jalabert2023,elalaily2024,ruf2024}.  The suggested microscopic mechanism is sketched in Fig.~\ref{fig3}{\bf c}: the NEPs, generated at the location of highest leakage current density (\ie, below the gate apex), form a local hot spot.  This hot spot is similar to that observed in superconducting single photon detectors, however, it is not caused by Joule heating~\cite{goltsman2001}.  When impinging on the superconductor, the NEPs excite quasiparticles by breaking up Cooper pairs, leading to a suppression of superconductivity.  While diffusing, both types of particles also undergo thermal relaxation, eventually limiting the radius of the suppression.

To quantify this hypothesis, we model the hot spot using a steady-state diffusion equation~\cite{supplementary},
\begin{align}
	D \nabla^2\rho(\vecr)+Q\delta(\vecr)-\frac{1}{\tau}\rho(\vecr) = 0 \ ,
\end{align}
adding a point-like source term, $Q\delta(\vecr)$, to account for excitation and a decay term, $\frac1\tau\rho(\vecr)$, to account for thermal relaxation.  Here,  $\rho(\vecr)$ is the density of particles at a distance $\vecr$ from the source, $D$ is the diffusion coefficient, $Q$ is the excitation rate, and $\tau$ is the thermal relaxation time.
Solving for $\rho$ yields the steady-state distribution of the particles,
\begin{align}
	\rho(r)
	&= \frac{\alpha \Pgate\tau}{4\pi \rd^2 r} \exp\left(-\frac{r}{\rd}\right) \ ,
	\label{eq:rho}
\end{align}
where $r=|\vecr|$ is the radial distance (assuming radial symmetry), $\rd = \sqrt{D\tau}$ is the diffusion length, and $Q=\alpha \Pgate$.  The parameter $\alpha$ describes the creation yield of NEPs.
From Eq.~(\ref{eq:rho}) we can deduce the power dependence of the hot-spot radius $\ro$.
For this, we assume a critical density $\rhoc =\rho(\ro)$ above which superconductivity is suppressed.  Solving for $\ro$, we find
\begin{align}
	\ro &= \rd W_0\left(\frac{e\Pgate}{\Pd}\right) \ ,
	\label{eq:ro}
\end{align}
where $W_0$ is the Lambert-$W$ function, and $\Pd = 4\pi e \rd^3\rhoc/(\alpha\tau)$ is the power for which $\ro=\rd$~\cite{supplementary}. Eq.~(\ref{eq:ro}) has only two free fit parameters, $\rd$ and $\Pd$.

Despite the simplicity of our model, which neglects the composite structure of the device, interface effects and nonequilibrium dynamics~\cite{xu2023}, we find excellent agreement between Eq.~(\ref{eq:ro}) and the experimental data.  Fit results are $\rd = 0.57\unit{\um}$ and $\Pd= 0.95\unit{\uW}$ for the dataset of device D3 (Fig.~\ref{fig3}{\bf a}), and $\rd = 0.65\unit{\um}$ and $\Pd= 0.36\unit{\uW}$ for the dataset of device D2b (Fig.~S4).
From our experiment, it is not possible to determine whether $\rd$ is dominated by diffusion of NEPs in the substrate or of quasiparticles in the superconductor.
For quasiparticle diffusion, $D \sim 5\ee{-4}\unit{m^2/s}$~\cite{jalabert2023} which would imply a lifetime of $\tau = \rd^2/D \sim 0.7\unit{ns}$.  This is roughly ten times longer than the time for hot-spot formation reported in~\cite{jalabert2023} and within the order of magnitude of superconducting single photon detector recovery~\cite{yang2007}.
For substrate diffusion, $D \sim 1\ee{-6}\unit{m^2/s}$~(for glass \cite{salazar2003}) implying $\tau \sim 350\unit{ns}$, which appears unreasonably long~\cite{joint2024}.
A future experiment could aim at measuring $\tau$ directly using time-resolved magnetometry~\cite{cui2017,herb2025}.
Follow-up experiments may also investigate the role of the substrate~\cite{zhang2018}.  If there were a strong phonon contribution to diffusion of the NEPs, the use of a material with a high thermal diffusivity $D$, such as diamond or AlN, should increase $\rd \propto \sqrt{D}$ and $\Pd \propto \sqrt{D^3}$.  Likewise, the injection efficiency would be modified.  A proper theoretical model would further account for interface effects and the exact phonon density of states by solving the Boltzmann transport equation (BTE)~\cite{kaviany2014, xu2023, chen2021}.

\subsection*{Dependence on gate geometry}

\begin{figure*}[ht]
	\centering
	\includegraphics[width=0.99\textwidth]{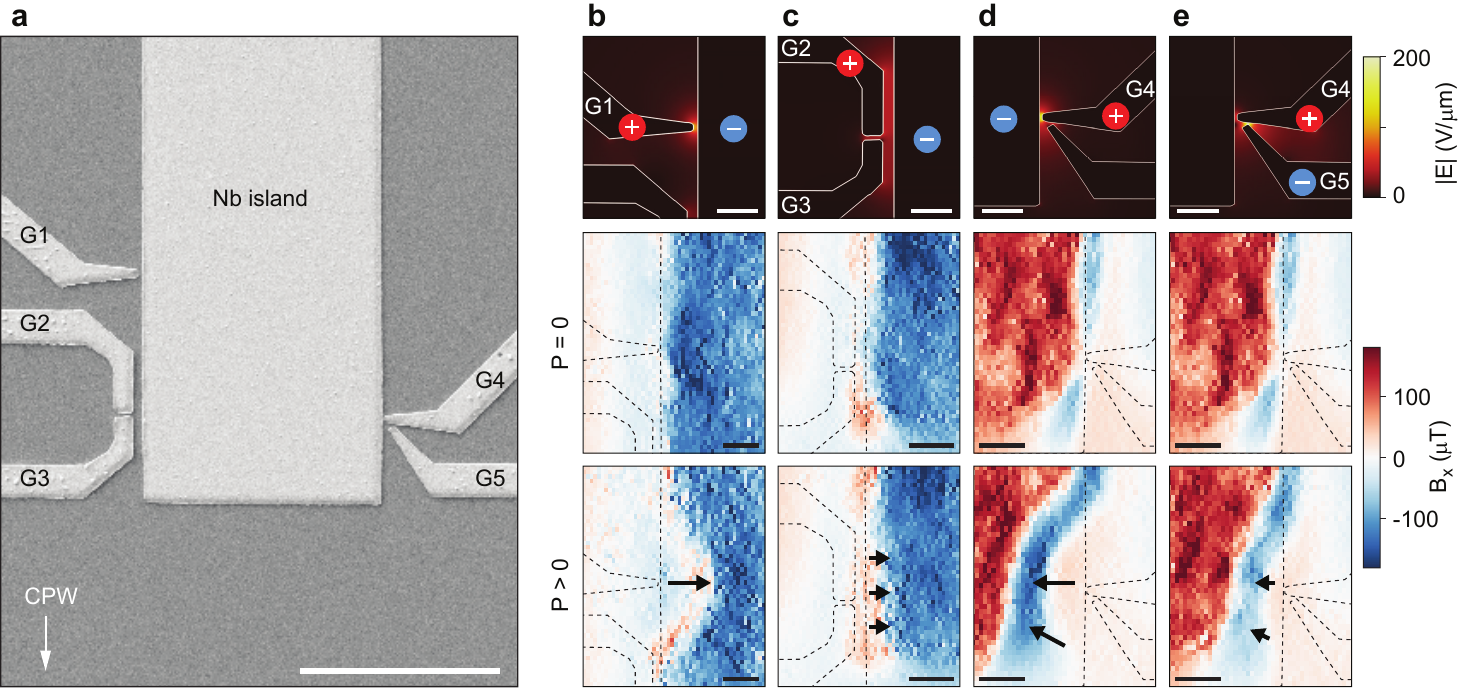}
	\caption{
		\textbf{Suppression of superconductivity for various gate geometries.}
		{\bf a}, Scanning electron micrograph of device D3. Scale bar, $5\unit{\um}$.
		{\bf b-e}, Electrode configurations. 
		The top row shows a FEM simulation of the electric field strength at $\Vgate=10\unit{V}$. The middle row shows the magnetometry map ($B_x$) for zero gate voltage ($\Pgate=0$). 
		The bottom row shows the corresponding magnetometry map for non-zero gate power ($5\unit{\uW}$ in {\bf b}, $2\unit{\uW}$ in {\bf c-e}).
		Dotted contours indicate approximate boundaries of the Nb island and gates (horizontally shifted to align with the stray field edge for clarity). Arrows indicate suppression with applied power.
		Note that the stray field is incompletely screened by the superconductor, presumably due to the presence of vortices and non-superconducting grains.  This leads to regions of negative $\Bx$ (blue in {\bf b,c}) or positive $\Bx$ (red in {\bf d,e}) above the Nb island, respectively.  See Fig.~S5 for an overview scan of the entire island.  
		Scale bars, $1\unit{\um}$.
	}
	\label{fig4}
\end{figure*}

Thus far, we have investigated the traditional side-gate geometry where the finger gate is within $100\unit{nm}$ from the large superconducting island.  In Fig.~\ref{fig4}, we extend the study to different gate geometries including gates at larger distances and extended gate electrodes (Fig.~\ref{fig4}{\bf a}). Transport experiments with similar geometries have indicated a GCS effect even for these situations~\cite{ritter2022}.

Fig.~\ref{fig4}{\bf b} shows quenching for the original finger-gate geometry (G1), placed far from the upper and lower corners of the device to eliminate boundary effects.  The electric field -- and hence the current density -- is concentrated in the region between the gate apex and island, effectively forming a point-like source on the scale of the observed quenching.  The suppression takes a circular shape at all imaged gate powers $\Pgate$ (Fig.~S6).  This justifies our earlier assumption of a point-like injection of NEPs and a radially symmetric hot spot.
 
Next, we investigate the extended gate electrode (G2).  Here, quenching is barely noticeable (Fig.~\ref{fig4}{\bf c}) and is elongated along the length of gates (G2) and (G3).  This makes intuitive sense: because the electric field extends along the $\sim 2.6\unit{\um}$ length of the electrodes (G2) and (G3), the current density is lower for the same gate power.  This results in an elongated quenching region and a shorter quenching distance.

Fig.~\ref{fig4}{\bf d} shows GCS quenching for a finger gate (G4) close to the lower right corner of the device.  Here, a distinct ``corner effect'' becomes apparent. (A similar effect is visible for higher powers in Fig.~\ref{fig2}{\bf c}).  The corner effect can be explained by the relatively long penetration depth that smoothens sharp features.  Finally, in Fig.~\ref{fig4}{\bf e}, we apply a voltage between (G4)-(G5) with the Nb island floating.  Strikingly, and in agreement with Ref.~\cite{ritter2022}, quenching occurs even in this situation.  However, the quenched area is reduced with respect to Fig.~\ref{fig4}{\bf d}, consistent with a hot spot that is removed by $\sim 0.8\unit{\um}$ from the island's edge compared to Fig.~\ref{fig4}{\bf d}~\cite{supplementary}.  Again, the observation is compatible with a point-like injection and subsequent diffusion of NEPs, where the effective distance to the injection site governs the extent of quenching.

\subsection*{Conclusions and Outlook}

Our experiments show a reproducible suppression of the Meissner screening in Nb microstructures under the application of power to an adjacent gate.  The range of full suppression is larger than $1\um$ and exceeds previous studies of nanowires by at least a factor of two~\cite{ruf2024}.  Required gate powers are $\sim\unit{\uW}$ and several orders of magnitude higher compared to transport measurements~\cite{ruf2024}.  This result is intriguing, and while it can be partially explained by the extended size of our Nb microstructure, it may hint at differences in the GCS in transport versus the GCS related to Meissner screening.

Future studies will investigate the interaction between superconducting vortices and the GCS.  Interestingly, as visible in Fig.~\ref{fig2}{\bf c}, the vortices completely escape the microstructure already at low gate voltages that are more typical for transport GCS.  Moreover, while all measurements show interactions between vortices and the GCS, the escape power is higher in the dataset acquired at $3.5\unit{K}$ (Fig.~S7) compared to the ones at $0.45\unit{K}$ (Fig.~\ref{fig2}{\bf c}) and $5.8\unit{K}$ (Fig.~S7).  The reasons for such GCS-vortex interactions are unknown and motivate further study.  Previous work on superconductors showed the movement of vortices due to applied current~\cite{embon2017} or magnetic force~\cite{timmermans2014, auslaender2009}, and the attraction of vortices to thermal hot spots~\cite{veshchunov2016} or regions with an enlarged quasiparticle population~\cite{engel2013}.

Finally, our experiments highlight some genuine advantages of scanning magnetic imaging techniques.  They enable investigation of the GCS effect with nanoscale resolution and over a temperature range from well below the superconducting gap to above the critical temperature of Nb.  Future studies may further correlate the magnetic images with electric field maps~\cite{huxter2023}, noise measurements~\cite{liu2025}, thermal mapping~\cite{laraoui2015,halbertal2016} or time-resolved sensing~\cite{herb2025}.


\vspace{0.5cm}\textbf{Acknowledgments}

We acknowledge support from the European Research Council through ERC CoG 817720 (IMAGINE) and the EU’s Horizon 2020 research and innovation programme under Grant Agreement no. 964398 (SUPERGATE).
We thank G.~Benga and Dr.~J.~Rhensius for taking the scanning electron micrographs, N.~Meinhardt for wirebonding the sample, Dr.~U.~Grob for maintenance and upgrade of the magnetometer, and A.~Zuschlag and M.~Hagner of the nano.lab at U.~Konstanz for technical support. We thank Prof.~H.J.M.~Schuttelaars, Prof.~G.~Steele and Prof.~Raffi Budakian for helpful discussions.

\vspace{0.5cm}\textbf{Author contributions}

S.D. initiated the collaboration.
S.D., P.J.S. and C.L.D. designed the project with input from E.S and A.D.B.
L.R. fabricated the devices and advised on device geometry and transport measurements.
P.J.S., K.J.K. and U.O. carried out the experiments and performed the data analysis.
P.J.S. and C.L.D. wrote the manuscript with the help of all authors.
All authors discussed the results.


\vspace{0.5cm}\textbf{Competing interests}

The authors declare no competing interests.

\vspace{0.5cm}\textbf{Data availability statement}

The data that support the findings of this study are available from the corresponding author upon reasonable request.

\vspace{0.5cm}\textbf{Additional information}

Supplementary information accompanies this paper. Correspondence and requests for materials should be addressed to C.L.D.


\vspace{0.5cm}\textbf{METHODS}\vspace{0.5cm}

{\bf Device fabrication}

Nb devices (thickness $\sim 20\unit{nm}$) with a Ti ($\sim 5\unit{nm}$) adhesion layer were patterned on a polished and chemical-cleaned $350\unit{\um}$-thick (100) Si substrate with a 300\,nm wet/dry/wet SiO$_2$ top layer (MicroChemicals). A $\sim 225\unit{nm}$ layer of poly(methyl methacrylate) (PMMA, Kayaku 950 A4) was spin-coated onto the substrate and baked on a hot plate at $180\oC$ for $90\unit{s}$. Full waveguide and device patterning was performed in a single step using electron beam lithography (20\,kV, $\sim 300\unit{\mu C/ cm^2}$ dose). The positive resist was developed in a solution of methyl isobutyl ketone (MIBK) and isopropanol (IPA). Nb and Ti films were deposited using RF magnetron sputtering (Ar flow: 17\,sccm, pressure: 1.5\,mTorr). Ti was deposited using a single RF gun at 200\,W with a deposition rate of $\sim 0.05\unit{nm/s}$. Nb was deposited using two RF guns and one DC gun, achieving a rate of $\sim 0.4\unit{nm/s}$. After deposition, the devices were immersed in a $50\oC$ acetone bath for several hours, followed by brief ultrasonication (a few seconds) for lifting off the mask. They were then rinsed with IPA and dried in an N$_2$ stream.

{\bf Electrical measurements}

The devices' electric properties were characterized using a Keithley 2450 source measure unit (SMU). The SMU was used in either a constant-voltage or constant-current mode setting.  The $\Ileak/\Vgate$ characteristics shown in Fig.~\ref{fig1}\textbf{c} were acquired in constant voltage mode, in order to enforce a uniform sampling of the voltage range.  By contrast, all magnetometry scans used the constant-current mode setting.  In addition, we implemented a constant-power mode on top of the constant-current mode using a software PID control loop.  In this constant-power mode, the gate voltage was continuously measured and the SMU current setpoint adjusted such that the power $\Pgate = \Ileak\Vgate$ remained constant.  This software constant-power mode was used in the measurements on device D3.

{\bf Leakage currents}

Since we measured the current and voltage at the source, the measured current include both the sample leakage and contributions from the wiring between the voltage source and the device. At low bias voltages ($|\Vgate| \lesssim 4\unit{V}$), the current scaled linearly with voltage, corresponding to an ohmic resistance of $\sim 60\unit{G\Omega}$, which we attributed to parasitic leakage in the wiring or connectors~\cite{ritter2021}. At higher gate voltages, the current increased exponentially and eventually dominated over the wiring contribution.

The leakage currents measured in our experiments showed a behavior similar to that reported from variable stress-induced leakage currents (vSILCs) \cite{elalaily2024,ruf2024}.  Firstly, the onset voltages observed in the $\Ileak/\Vgate$ characteristics varied from measurement to measurement, as shown in Fig.~\ref{fig1}{\bf c}.  Moreover, we observed that the gate voltage (or gate power) varied over the course of hours while operating the SMU in constant current mode (Fig.~S2).
Other possible origins of $\Vgate$ variations include charge accumulation and capture in the gate dielectric giving an additional capacitance, involvement of surface traps that are sensitive to the enviroment (\eg, air exposure), and the onset of filament formation due to high electrical stress~\cite{yao2010}.

{\bf Magnetometry measurements}

The magnetometry images were acquired using modified atomic force/confocal microscope from Attocube, integrated into the cold insertable probe of a Leiden Cryogenics CF-CS100 dry dilution refrigerator. Details of the setup can be found in Refs.~\cite{scheidegger2022,scheidegger2024thesis}. For scanning, we used a commercial NV scanning probe from QZabre.  The tuning fork was operated in non-contact atomic force microscopy (AFM) mode with a NV-to-sample separation of approximately $\zo\sim 100\unit{nm}$.  Measurements were performed using several NV probes, all cut along the $\langle100\rangle$ crystal orientation.

Static out-of-plane magnetic fields were applied using a superconducting vector magnet (American Magnetics Inc.).  The projection of the magnetic field along the NV quantization axis $B_\mathrm{NV}$ was measured by determining the frequency of the $m_\mathrm{s} = 0$ to $m_\mathrm{s} = -1$ spin transition, $f_{0,-1}$, using the pulsed ODMR technique~\cite{dreau2011}. 
$B_\mathrm{NV}$ was obtained by comparing the resonance frequency $f_{0,-1}$ to the zero-field resonance $f_\mathrm{ZFS}$ via the relation $B_\mathrm{NV} = 2\pi(f_\mathrm{ZFS}-f_{0,-1})/\ye$, where $\ye/2\pi=28.025\unit{MHz/mT}$ is the gyromagnetic ratio of the electron.  The projection axis of $\BNV$ is given by the crystallographic orientation of the single crystal diamond tip, and is approximately $\theta = 54^\circ$ and $\phi = \{0^\circ,90^\circ,180^\circ,270^\circ\}$, where $\theta$ and $\phi$ are the polar and azimuth angle in the scanning frame of reference.
Typical magnetic field sensitivities at cryogenic temperatures were between $\sim10-20\unit{\mu T/ \sqrt{Hz}}$.  Finally, we applied a Fourier transform-based algorithm~\cite{lima2009} to reconstruct the three-component vector field $\mathbf{B} = \left(B_x, B_y, B_z\right)$ from the measured $\BNV$ projection.

{\bf Magnetic field simulations}

Simulations of the magnetic stray field were carried out using the SuperScreen software package \cite{bishopvanhorn2022}. 
For this, we assumed a device geometry of uniform thickness $d\ll\lambda$, allowing us to model the geometry as a two-dimensional problem with a sheet supercurrent density and an effective penetration depth $\lambda _\mr{eff} = \lambda^2/d$.  The geometry of the layer was then partitioned to perform Finite Element (FEM) simulations.  The magnetic response was then obtained by setting a bias field and modeling the supercurrent density via the stream function~\cite{brandt2005}.  The simulations in Fig.~\ref{fig2}{\bf d} used a rectangular island while cutting out a circular section of radius $\ro$ at the gate-facing side of the island.


\input{"references.bbl"}

\end{document}

%% file: references.bbl
%